\documentclass[pra,twocolumn,reprint,superscriptaddress,amsmath,amssymb,
aps,showpacs, floatfix]{revtex4}

\usepackage{graphicx}
\usepackage{dcolumn}
\usepackage{bm}
\usepackage{epstopdf}

\newcommand{\Op}[1]{\boldsymbol{\mathsf{\hat{#1}}}}

\newcommand{\Fkt}[1]{\,\mathsf {#1}}
\newcommand{\eps}{\varepsilon}
\newcommand{\vphi}{\varphi}

\def\openone{\leavevmode\hbox{\small1\kern-3.3pt\normalsize1}}

\begin{document}

\title{Optimizing entangling quantum gates for physical systems}

\author{M. M. M\"uller}
\affiliation{Institut f\"ur Quanteninformationsverarbeitung, 
Universit\"at Ulm, 89081 Ulm, Germany}

\author{D. M. Reich}
\affiliation{Institut f\"ur Theoretische Physik,
  Freie Universit\"at Berlin,
  Arnimallee 14, 14195 Berlin, Germany}
\affiliation{Institut f\"ur Physik,
  Universit\"at Kassel,
  Heinrich-Plett-Str. 40, 34132 Kassel, Germany}

\author{M. Murphy}
\affiliation{Institut f\"ur Quanteninformationsverarbeitung, 
Universit\"at Ulm, 89081 Ulm, Germany}

\author{H. Yuan}
\affiliation{Research Laboratory of Electronics, 
Massachusetts Institute of Technology, Cambridge, MA 02139}

\author{J. Vala}
\affiliation{Department of Mathematical Physics, National University
  of Ireland, Maynooth, Ireland} 
\affiliation{School of Theoretical Physics, Dublin Institute for
  Advanced Studies, 10 Burlington Rd., Dublin, Ireland} 

\author{K. B. Whaley}
\affiliation{Department of Chemistry, 
University of California, Berkeley, California 94720, USA}

\author{T. Calarco}
\affiliation{Institut f\"ur Quanteninformationsverarbeitung, 
Universit\"at Ulm, 89081 Ulm, Germany}

\author{C. P. Koch}
\affiliation{Institut f\"ur Theoretische Physik,
  Freie Universit\"at Berlin,
  Arnimallee 14, 14195 Berlin, Germany}
\affiliation{Institut f\"ur Physik,
  Universit\"at Kassel,
  Heinrich-Plett-Str. 40, 34132 Kassel, Germany}
\email{christiane.koch@uni-kassel.de}

\date{\today}

\begin{abstract}
 Optimal control theory is a versatile tool that presents a route to
 significantly improving figures of merit for quantum information
 tasks.   
We combine it here with the geometric theory for local equivalence classes
 of two-qubit operations to derive an optimization algorithm that
 determines the best entangling two-qubit gate for a given physical
 setting. We demonstrate the power of this approach for trapped
 polar molecules and neutral atoms. 
\end{abstract}

\maketitle

\section{Introduction}
Quantum information science requires extreme accuracy in
implementation of quantum tasks
such as gate operations, in order to fulfill the stringent
criteria of fault tolerance \cite{NielsenChuang}. However the dynamics
underlying a gate operation 
is often very complex, occurring in a Hilbert space much larger
than that of the qubits.
Optimal quantum control then provides an indispensable tool
for obtaining a high-fidelity implementation \cite{PalaoPRA03}. 
Realization of a universal set of gates
comprises multiple levels of difficulty, because different physics is typically involved 
for one- and two-qubit operations.  For many qubit systems,  
two-qubit gates are the most challenging since they involve controlled 
interaction between two otherwise isolated quantum systems. 
Furthermore, for a given physical implementation, it is not
necessarily \textit{a priori} clear which two-qubit gate can best be
implemented 
when the practical criteria of optimal achievable fidelity and
realistic gate operation time are imposed. 

We address this problem here by combining optimal control theory 
\cite{PalaoPRA03} with the geometric theory for
local equivalence classes of two-qubit operations
\cite{ZhangPRA03}.This allows us to develop an
algorithm that optimizes for the 
non-local content of a two-qubit gate rather than for fidelity of a specific gate such
as CNOT.
The resulting separation of non-local from local control
objectives relaxes the control constraints and 
enables both maximum fidelities to be reached and fundamental limits
for control 
to be identified. 
We apply our algorithm to
trapped polar molecules and neutral atoms, both 
candidates for realizing quantum computation.  
Manipulation of trapped polar molecules with microwave
fields has been shown to allow realization of effective spin-spin
models \cite{MicheliNatPhys06} 
with continuously tunable parameters.  
In our first example we use the combined
optimal/geometric control algorithm to determine which 
non-local gates can be realized for a given underlying
molecule-microwave field Hamiltonian.  
In our second example, we obtain optimal solutions for a Rydberg gate
\cite{JakschPRL00} between trapped neutral atoms 
in the presence of both decay processes and entangling couplings
between internal and external degrees of 
freedom. We show that high fidelity Rydberg gate operation is
possible even for a configuration where the blockade regime is not reached and
despite spontaneous emission from intermediate states.  

\section{Optimizing the non-local content of a two-qubit gate}

\subsection{Optimal control theory}
\label{subsec:oct}
High-fidelity implementations of quantum gates can be
obtained with optimal control theory by defining a suitable distance
measure between the desired unitary $\Op O$ and the actual evolution,
e.g.,
\begin{equation}
  \label{eq:J_T}
  J^{D}_T=1-\frac{1}{N} \mathfrak{Re}\left[\Fkt{Tr}\left\{
      \Op{O}^+\Op{P}_N\Op{U}(T,0;\eps)\Op{P}_N
    \right\}\right]\,,
\end{equation}
and minimizing it with respect to some external field $\eps(t)$~\cite{PalaoPRA03}.
Here, $\Op{U}(T,0;\eps)$  represents the evolution of the system
under the action of an external field  from time $0$ to time $T$. 
For example, $\eps(t)$ can be a pulsed laser field or a time-dependent
magnetic field. The Hilbert space in which the system evolves is
possibly very large. The logical subspace, i.e., the 
subspace of the total Hilbert space in which $\Op{O}$ acts has
dimension $N$, $N=\Fkt{dim}\,\mathcal H_O$
($N=4$ for a two-qubit gate). $\Op{P}_N$ denotes the projector onto
this subspace. The trace is evaluated by choosing a suitable
orthonormal basis of the subspace $\mathcal H_O$,
$\{|\vphi_{k=1,\ldots,N}\rangle\}$. The evolution 
of the system is thus expressed in terms of the time evolved basis
states, $|\vphi_k(T)\rangle=\Op U(T,0;\eps)|\vphi_k\rangle$, and $J_T^D$
becomes a functional of the states $|\vphi_k(T)\rangle$.
For the specific choice of Eq.~\eqref{eq:J_T}, 
$J^{D}_T$ is a phase sensitive functional that depends linearly on the
states, $|\vphi_k(T)\rangle$, and  equals zero for
perfect implementations of $\Op{U}$. 

Additional constraints can be introduced to ensure finite pulse fluence,
\begin{equation}
  \label{eq:g_a}
  g_a=\lambda_a\int_0^T \left[\eps(t)-\eps_{ref}(t)\right]^2/S(t) dt\,,   
\end{equation}
or avoid population of states subject to loss,
\begin{equation}
  \label{eq:g_b}
  g_b=\frac{\lambda_b}{NT}\int_0^T\sum_{m=1}^N
  \langle \varphi_m(t)|\Op P_{avoid}|\varphi_m(t)\rangle dt\,.
\end{equation}
Here, $\eps_{ref}(t)$ denotes a reference field, $S(t)$ is a shape
function to switch the field smoothly on and off, and $\lambda_a$ and
$\lambda_b$ are weights \footnote{
  $\lambda_a$ and $\lambda_b$ are numerical parameters of the
  algorithm:
  The choice of $\lambda_a$ determines the magnitude of the change in
  the field, i.e., the step size,
  and thus convergence speed~\cite{PalaoPRA03}. The ratio of
  $\lambda_a$ and $\lambda_b$ controls the relative weight of the two
  constraints. If $\lambda_b \ll \lambda_a$, $g_b$ plays only a minor
  role, if $\lambda_b \gg \lambda_a$, $g_b$ is strictly enforced. The
  latter typically comes at the expense of slower convergence since
  optimization under an additional constraint represents a more
  difficult control problem~\cite{PalaoPRA08}.
}. 
 $\Op P_{avoid}$ denotes 
the projector onto the subspace of the total Hilbert space that shall
never be populated, and $g_b$ minimizes population of this subspace
\cite{PalaoPRA08}.   
The total functional to be minimized, $J$, is given by the sum of
$J_T^D$, $g_a$ and 
$g_b$,
\begin{equation}
  \label{eq:J}
  J = J_T^D + g_a + g_b\,.
\end{equation}
Solving Eq.~\eqref{eq:J} with an iterative procedure in which the
reference field, $\eps_{ref}(t)$, is chosen at each level of iteration
to be the optimized field from the 
previous iteration ensures vanishing of $g_a$ at the optimum, since this is
reached when the optimal value of $J$ is determined only by
$J_T^D$ and possibly $g_b$, but not by the pulse fluence
\cite{PalaoPRA03}. 
A monotonically convergent algorithm is obtained for this control problem
using a simplified version of Krotov's method
\cite{PalaoPRA03,PalaoPRA08}. 

The core idea of Krotov's method~\cite{Konnov99,SklarzPRA02} 
consists in disentangling the interdependence of the states and the
field. This is achieved by separating the final-time and intermediate-time
dependencies of the total functional $J$ and 
adding to this a vanishing quantity, $\Phi_T-\Phi_0 -\int_0^T \dot{\Phi} dt$,
which is expressed in terms of a functional $\Phi$ 
that depends only on the states and not on the field. 
The freedom of choice in $\Phi$ is utilized to ensure monotonic
convergence of the algorithm.  Specifically, expanding $\Phi$ up to second order in
the states, $\{\vphi_k(t)\}$, the expansion coefficients are chosen
such that the first and second order derivatives of $J$ with respect
to the states fulfill the conditions for the extremum and
maximum of $J$, respectively, when $J$ is minimized. 
Since $J$ thus takes the worst possible value with respect to the
choice of basis states, any change of $J$ due to varying the  
field, $\eps(t)$, then leads to improvement toward the actual target,
minimization of $J$. 

For the simple functional of Eq.~\eqref{eq:J}, it
turns out that the second order conditions are trivially fulfilled by
correct choice of the sign of the weights $\lambda_a$ and $\lambda_b$,
and the second order expansion coefficients of $\Phi$ can be set to zero
\cite{PalaoPRA03}. The ensuing algorithm thus coincides with that
obtained from straightforward variation of $J$ with a specific
discretization of the coupled control
equations~\cite{ZhuJCP98,MadayJCP03}.  
It yields pulses that implement the
desired gate with high fidelity, provided
that  the dynamics allow for it
\cite{Goerz}. For functionals with higher than quadratic dependence
on the states, however, 
it is essential to include the second order contribution to $\Phi$ in
order to ensure monotonic convergence. The second order expansion 
coefficient can be estimated 
either analytically or numerically, as detailed in Ref.~\cite{Reich10}.

\subsection{Geometric theory of non-local two-qubit operations}
\label{subsec:geo}
The group of all two-qubit gates, $SU(4)$, consists of local
operations, $SU(2)\otimes SU(2)$, and non-local 
operations, $SU(4)\backslash SU(2)\otimes SU(2)$. This is a direct
result of the existence of a Cartan decomposition of the
corresponding Lie algebra \cite{Helgason78}. Any two-qubit operation can be
written as 
\begin{equation}
  \label{eq:U}
  \Op{U}=\Op{k}_1\Op A \Op{k}_2  
\end{equation}
where 
\begin{equation}
  \label{eq:A}
  \Op A = \exp\left[-\frac{i}{2}\sum_{j=x,y,z}c_j
  \Op{\sigma_j}\otimes\Op{\sigma_j}\right]  
\end{equation}
and the $\Op{k}_n \in SU(2)\otimes SU(2)$ are
local operators. The set $SU(4)\backslash SU(2)\otimes SU(2)$ 
is generated by the maximal Abelian
subalgebra of the $su(4)$ algebra which 
is spanned by the operators $\Op{\sigma_j}\otimes\Op{\sigma_j}$, $j = x, y , z$. 
Since these two-qubit operators commute, the operations belonging to
$SU(4)\backslash SU(2)\otimes SU(2)$ can be represented 
by three real numbers $(c_x,c_y,c_z)$. Due to the periodicity of the
complex exponential, the $c_j$ take their value from a
three-dimensional cube $\mathbf{I}^3$ with edges $\mathbf{I} = [0, \pi]$.  
The operations from this set create and change entanglement between
two qubits (with the exception of the identity operation $\openone$ and
the SWAP gate for which all $c_j = 0 \mod\pi$ and all $c_j = \pi/2$,
respectively). 
Each local operation $\Op{k}_{1/2} \in SU(2)\otimes SU(2)$ 
accounts for additional 6 parameters, yielding the remaining 
12 parameters that 
are required to fully characterize the elements of $SU(4)$.  

\begin{figure}[tb]
  \centering
  \includegraphics[width=0.75\linewidth]{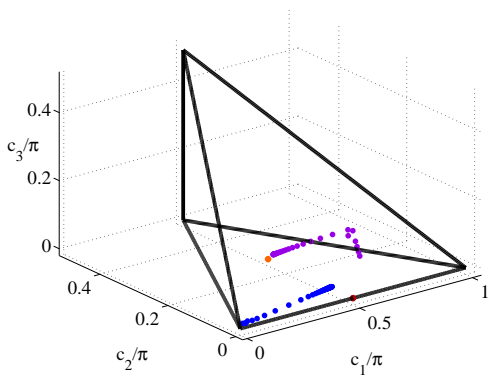}
  \caption{(Color online) 
    Optimization of non-local gates in the Weyl chamber.  
    Local invariants of the B gate (orange) and 
    CNOT (red) are approached iteratively (each violet/blue
    point corresponds to one step in the optimization of the
    effective spin-spin Hamiltonian, cf. Section~\ref{subsec:toy}
    below). 
}
  \label{fig:weyl}
\end{figure}

The representation of the non-local content of $\Op{U}$ in terms of
the $c_j$ is not unique. Different points in the cube $\mathbf{I}^3$ may
correspond to the same two-qubit operation up to local
transformations \cite{ZhangPRA03}.  
For example, all eight corners of the cube are equivalent to the
identity operator up to local transformations. This symmetry is
characterized by the Weyl reflection group. It is generated  by
permutations or permutations with  
sign flips of two entries in $(c_x,c_y,c_z)$. Symmetry reduction  of
the cube $\mathbf{I}^3$ leads to a geometric representation of non-local
two-qubit gates 
within the Weyl chamber $a^+$ which is the tetrahedral
segment of the cube  $\mathbf{I}^3$  spanned by  
$(c_x,c_y,c_z)=(0,0,0)$, $(\pi,0,0)$, $(\pi/2,\pi/2,0)$, and 
$(\pi/2,\pi/2,\pi/2)$. All two-qubit gates that are equivalent up to
local operations $\Op{k}_n$ are geometrically represented by a single 
point in the Weyl chamber (except on its base where local equivalence
classes may be represented by two symmetry-equivalent points)
\cite{ZhangPRA03}.  
For example, CNOT and the controlled $\pi$-phase gate (CPHASE) are in
the same local equivalence class, which is 
defined by the point $(\pi/2,0,0)$, cf. Fig.~\ref{fig:weyl} (left
panel).  

Each class of all the 
two-qubit gates that are equivalent up to local operations,
$[\Op{U}]$, can also be 
characterized by three real numbers~\cite{MakhlinQIP02}, the 
local invariants $g_1$, $g_2$ and $g_3$. 

It was shown in Ref.~\cite{ZhangPRA03} that there
exists a one-to-one correspondence between the points in the Weyl
chamber $(c_x,c_y,c_z)$ and the local
invariants. For a given unitary $\Op U$, $g_1$,
$g_2$, $g_3$ are easily calculated from the representation of $\Op U$
in the Bell basis ~\cite{ZhangPRA03}.
Denoting the transformation from the logical basis into the Bell basis
by $\Op Q$, $\Op{U}_B=\Op{Q}^+\Op{U}\Op{Q} = \Op{o}_1\Op{F} \Op{o}_2$ where
$\Op{o}_n = \Op{Q}^+\Op{k}_n\Op{Q} \in SO(4)$ and thus
$\Op{o}_n^T\Op{o}_n = \Op{o}_n\Op{o}_n^T = \openone$. 
The matrix $\Op{F}$ is diagonal, 
\begin{eqnarray}
  \label{eq:F}
  \Op F &=&\Fkt{diag}\{e^{\frac{i}{2}(c_x - c_y + c_z)}, 
  e^{\frac{i}{2}(c_x + c_y - c_z)}, \nonumber \\
  && \quad\quad e^{-\frac{i}{2}(c_x + c_y + c_z)},
  e^{\frac{i}{2}(-c_x + c_y + c_z)}\}\,.  
\end{eqnarray}
Introducing a matrix $\Op{m}_U$,
\begin{equation}
  \label{eq:m}
  \Op m_U= \Op{U}_B^T \Op{U}_B = \Op{o}_2 \Op{F}^2 \Op{o}_2\,,  
\end{equation}
leads to automatic elimination of the first local factor $\Op{o}_1$.
The local invariants are defined~\cite{ZhangPRA03} as
\begin{subequations}\label{eq:gs1}
\begin{eqnarray}
  g_1 &=& \frac{1}{16}\mathfrak{Re} \Fkt{Tr}\{\Op{m}_U\}^2   \\
  & = &     \cos^2c_x \cos^2c_y \cos^2c_z 
    - \sin^2c_x \sin^2c_y \sin^2c_z\,, \nonumber \\
  g_2 &=& \frac{1}{16}\mathfrak{Im} \Fkt{Tr}\{\Op{m}_U\}^2 \\
  & = & \frac{1}{4} \sin2c_x \sin2c_y \sin2c_z\,, \nonumber \\
  g_3 &=& \frac{1}{4}\Fkt{Tr}\{\Op{m}_U\}^2 - \Fkt{Tr}\{\Op{m}_U^2\} \\
  & = & 4 \cos^2c_x \cos^2c_y \cos^2c_z 
  - 4\sin^2c_x \sin^2c_y \sin^2c_z \nonumber \\
  && - \cos2c_x \cos2c_y \cos2c_z\,, \nonumber
\end{eqnarray}  
\end{subequations}
where the remaining local factor $\Op{o}_2$ is eliminated due to the 
cyclic permutation invariance of the trace.  
To generalize the local invariants to the elements of the group
$U(4)$, i.e., $\Op{U} = e^{i\alpha} \Op{U}' \in U(4)$ where $\Op{U}'
\in SU(4)$, the global phase $e^{i\alpha}$ is eliminated by dividing
the local invariants by $\det\{\Op{U}\} = e^{4i\alpha}$. The final form
of the local invariants is then 
\begin{subequations}\label{eq:gs}
\begin{eqnarray}
  g_1 &=& \mathfrak{Re} \Fkt{Tr}\{\Op{m}_U\}^2 / 16 \Fkt{det}\{\Op{U}\}\,, \\
  g_2 &=& \mathfrak{Im} \Fkt{Tr}\{\Op{m}_U\}^2/ 16 \Fkt{det}\{\Op{U}\}\,,\\
  g_3 &=& \Fkt{Tr}\{\Op{m}_U\}^2 - \Fkt{Tr}\{\Op{m}_U^2\}/ 4 \Fkt{det}\{\Op{U}\}\,.
\end{eqnarray}  
\end{subequations}
A few examples of local equivalence classes, their coordinates
$c_i$ in the
Weyl chamber and the corresponding Makhlin invariants $g_i$ are given in 
Table~\ref{tab:LEC}.
\begin{table}[tb]
  \centering
  \begin{tabular}{ | c || c | c | c || c | c | c | }
    \hline
    class & ~~$c_x$~~ & ~~$c_y$~~ & ~~$c_z$~~ & 
    ~~$g_1$~~ & ~~$g_2$~~ & ~~$g_3$~~ \\ \hline\hline
    [$\openone$] & $0$ & $0$ & $0$ & $1$ & $0$ & $3$ \\ \hline
    [CNOT] & $\pi/2$ & $0$ & $0$ & $0$ & $0$ & $1$ \\ \hline
    [CPHASE] & $\pi/2$ & $0$ & $0$ & $0$ & $0$ & $1$ \\ \hline
    [B-gate] & $\pi/2$ & $\pi/4$ & $0$ & $0$ & $0$ & $0$ \\ \hline
    [$\sqrt{\mathrm{SWAP}}$] & $\pi/4$ & $\pi/4$ & $\pi/4$ & $0$ & $1/4$ & $0$ \\ \hline
    [SWAP] & $\pi/2$ & $\pi/2$ & $\pi/2$ & $-1$ & $0$ & $-3$ \\ 
    \hline
  \end{tabular}
  \caption{Examples of local equivalence classes, their coordinates
    $(c_x,c_y,c_z)$ in the Weyl chamber, and their local
    invariants $g_1$, $g_2$, $g_3$}.
  \label{tab:LEC}
\end{table}

\subsection{Optimization functional based on the local invariants}
\label{subsec:J_LI}
To act as a suitable optimization functional, 
any real-valued functional $J[\{\vphi_k\}]$
should fulfill two necessary conditions:
(i) $J$ must strictly take its global optimum for all sets of
states $\{\vphi_k\}$ that represent the desired outcome, and (ii) $J$
must be regular, i.e., at least twice differentiable. Moreover, as a
matter of practicality, it should be possible to express
$J[\{\vphi_k\}]$ explicitly in terms of the states, $\{\vphi_k\}$,
in order to carry out the differentiation. 
While the Weyl chamber coordinates, $(c_x,c_y,c_z)$, are only
implicitly given in terms of the evolution $\Op U$ of
the system, the local
invariants $g_1$, $g_2$, $g_3$ can directly be calculated from $\Op U$
by use of Eqs.~\eqref{eq:gs1}
(or Eqs.~\eqref{eq:gs} for $\Op U \in U(4)$), and hence from the
time-evolved basis states 
$|\vphi_k(t)\rangle= \Op U|\vphi_k(0)\rangle$.   
Therefore the local invariants lend themselves
naturally to the 
definition of a distance measure between the desired local equivalence
class $[\Op{O}]$ and the actually realized local equivalence
class $[\Op{U}]$. We define this distance measure as
\begin{equation}
  \label{eq:d}
  d=\sum_{i=1}^{3}\Delta g_i^2\quad \mathrm{with}\quad
  \Delta g_i=|g_i(\Op{O})-g_i(\Op{U}_{T,N})|  \,.
\end{equation}
The distance $d$ constitutes one component of the desired optimization functional.
Since the time evolution of the physical system typically occurs in a
Hilbert space that is much larger than the logical space, 
$\Op U(T,0;\eps)$ needs to be projected into the logical subspace, 
\begin{equation}
  \label{eq:U_TN}
 \Op U_{T,N} = \Op P_N \Op U(T,0;\eps) \Op P_N\,.
\end{equation}
Correspondingly, $d$ needs to be augmented by a term that enforces 
unitarity in the logical space at the final time $T$, leading to the following
optimization functional for a non-local equivalence class:
\begin{equation}
  \label{eq:Jnew}
  J_T^{LI} = \Delta g_1^2 + \Delta g_2^2 + \Delta g_3^2 + 1 -
  \frac{1}{N} \Fkt{Tr}\left\{\Op U_{T,N} \Op U_{T,N}^+\right\}\,.
\end{equation}
Using the definition of the local invariants, the $\Delta g_i$ are 
expressed in terms of the time-evolved basis states as follows. 
Expanding $\Op U$ in the orthonormal basis
$|\vphi_k\rangle$, the matrix elements of $\Op m_U$ are second order
in the states, cf. Eq.~\eqref{eq:m}. Therefore the $g_j$ are fourth
order in the states, 
cf. Eq.~\eqref{eq:gs1}, and 
$J^{LI}_T$ turns out to be a polynomial of eighth
order in the states.  The specific form of $J^{LI}_T$ is given in
Appendix~\ref{app:J}.  

For such a non-convex functional, the commonly used
optimization algorithms including that of Ref.~\cite{PalaoPRA03} are
not sufficient to ensure monotonic convergence and indeed fail to
converge. To the best of our knowledge, Krotov's method is the only
approach providing
an optimization algorithm that ensures monotonic convergence for
arbitrary functionals in quantum control~\cite{Reich10}. For difficult
control problems, monotonic convergence is essential to reach any
optimum, even a local one. Applying Krotov's method to the functional
$J_T^{LI}$, 
the optimal field is obtained iteratively according to
\begin{widetext}
\begin{eqnarray}
  \label{eq:field}
  \eps^{(i+1)}(t) =  \eps^{(i)}(t) + \frac{S(t)}{\lambda_a} \mathfrak{Im}\left\{
    \sum_{k=1}^N \left\langle \chi_k^{(i)}(t)\bigg|
    \frac{\partial \Op H}{\partial \eps}^{(i+1)} \bigg|
      \varphi_k^{(i+1)}(t)\right\rangle 
    + \frac{1}{2}\sigma(t)\sum_{k=1}^N \left \langle 
      \Delta\varphi_k^{(i+1)}(t) \bigg|
      \frac{\partial \Op H}{\partial \eps}^{(i+1)}
    \bigg|
    \varphi^{(i+1)}_k(t)\right\rangle 
  \right\}\,,
\end{eqnarray}
\end{widetext}
where
$|\Delta\varphi^{(i+1)}_k(t)\rangle=
|\varphi^{(i+1)}_k(t)\rangle-|\varphi^{(i)}_k(t)\rangle$.
The details of the iterative algorithm are presented in
Appendix~\ref{app:OCT}. In brief, 
Eq.~(\ref{eq:field}) implies propagating the basis states 
$|\varphi_k\rangle$ forward in time and the adjoint states
$|\chi_k\rangle$ backward in time with the 'initial' condition,
$|\chi_k(T)\rangle$, determined by the functional $J^{LI}_T$. 
$S(t)/\lambda_a$ and $\sigma(t)$ are parameters of the optimization
algorithm that 
are constructed following the prescription of Ref.~\cite{Reich10},
cf. Appendix~\ref{app:OCT}. 

\subsection{Actual gate operation and gate error}
\label{subsec:U}
Optimizing the functional $J^{LI}_T$ yields some gate $\Op{U}_{T,N}$
that is, up to some small error, in the local equivalence class of the
desired gate $\Op{O}$. 
In order to actually implement $\Op{O}$, we
need to determine the local operations $\Op{k}_1$ and $\Op{k}_2$ such
that $\Op{O}=\Op{k}_1\Op{U}\Op{k}_2$. 
This is achieved by transforming both $\Op{O}$ and
$\Op{U}_{T,N}$ to the canonical form  $\Op A$,  
$\Op k_1^\prime\Op O\Op k_2^\prime = \Op A$,  
$\Op k_1^{\prime\prime}\Op{U}_{T,N}\Op k_2^{\prime\prime} = \Op A$.
Since $\Op A$ is diagonal in the Bell
basis, $\Op k_1^\prime$, $\Op k_2^\prime$, $\Op k_1^{\prime\prime}$, 
$\Op k_2^{\prime\prime}$ are obtained by diagonalization, and their
combination yields $\Op{k}_1$, $\Op{k}_2$. 
We first determine the local operations $\Op k_1^\prime$, $\Op k_2^\prime$ that
transform $\Op U_{T,N}$ into the canonical form $\Op A$.
This is achieved by calculating
the $g_i(\Op U_{T,N})$ which yield the $c_i$ and thus $\Op A$. 
In the Bell basis, $\Op A$ is diagonal and $\Op k_1^\prime$ and $\Op
k_2^\prime$ are the transformations that diagonalize 
$\Op U_{T,N} \Op U_{T,N}^T$ and $\Op U_{T,N}^T\Op U_{T,N}$, each yielding
$\Op A^2$. 
Care must be taken to assure the same ordering of eigenvectors when
determining $\Op k_1^\prime$ and $\Op k_2^\prime$. 
Repeating the same procedure for the local
operations $\Op k_1^{\prime\prime}$, $\Op k_2^{\prime\prime}$ that transform the
target operation $\Op O$ into the canonical form $\Op A$, and combining 
$\Op k_1^\prime$, $\Op k_2^\prime$, $\Op k_1^{\prime\prime}$, $\Op
k_2^{\prime\prime}$ yields $\Op{k}_1$, $\Op{k}_2$. 
Assuming the errors
associated with the local operations $\Op{k}_1$, $\Op{k}_2$ to be
small compared to the error of the non-local operation, 
the actual gate error, $\mathcal{E}$, is then obtained by evaluating
$J_T^D$ for $\Op{k}_1\Op{U}_{T,N}\Op{k}_2$,
\[
\mathcal{E} = 1 - \frac{1}{N}\mathfrak{Re}\left[\Fkt{Tr}\left\{
      \Op{O}^+\Op{k}_1\Op{U}_{T,N}\Op{k}_2
    \right\}\right]\,.
\]

\section{Applications}
\label{sec:app}

We apply the local invariants optimization functional to two examples,
an effective 
spin-spin model that can be realized by trapped polar molecules and a
Rydberg gate for trapped atoms. In the first example, the Hamiltonian
may become complex, making it impossible to determine \textit{a
  priori} which two-qubit gates it can implement. The Hamiltonian of
the second example can realize diagonal
two-qubit gates only. The complexity that necessitates use of optimal
control theory in this case draws from coupling
the logical basis to external degrees of freedom. 

\subsection{Two-qubit gates for an effective spin-spin model}
\label{subsec:toy}
Trapped polar molecules with $^2\Sigma_{1/2}$ electronic ground
states, subject to near-resonant microwave driving inducing 
strong dipole-dipole coupling, give rise to the effective Hamiltonian 
\begin{equation}
  \label{eq:Heff}
\Op H_{eff}(t) = \frac{\hbar|\Omega(t)|}{8}\sum_{i,j=1}^4 \Op\sigma_i 
  A_{ij}(x_0,t) \Op\sigma_j
\end{equation}
within second-order perturbation theory in the field
\cite{MicheliNatPhys06} ($\Op\sigma_4=\openone$). The couplings 
$A_{ij}(x_0,t)=|\Omega(t)|a_{ij}(x_0)$ depend on the
distance $x_0$ between the molecules and on the polarization, 
detuning and possibly time-dependent envelope of the microwave field.
We consider here SrF molecules 
in an optical lattice with a lattice spacing of $300\,$nm and microwave
radiation of about $15\,$GHz.   
The qubit is represented by the spin of the valence electron of the
molecule in its rotational ground state, as described in
Ref. \cite{MicheliNatPhys06}.  
We seek to implement two-qubit gates that are locally 
equivalent to the B gate \cite{ZhangPRL04} and to CNOT, cf.
Fig.~\ref{fig:weyl} and Table~\ref{tab:LEC}.  
The B gate allows for generating a generic
two-qubit operation from just two successive applications. Arbitrary
two-qubit operations can thus be implemented with a minimal 
count of
two-qubit and single-qubit gates \cite{ZhangPRL04}.

\begin{figure}[tb]
  \centering
  \includegraphics[width=0.9\linewidth]{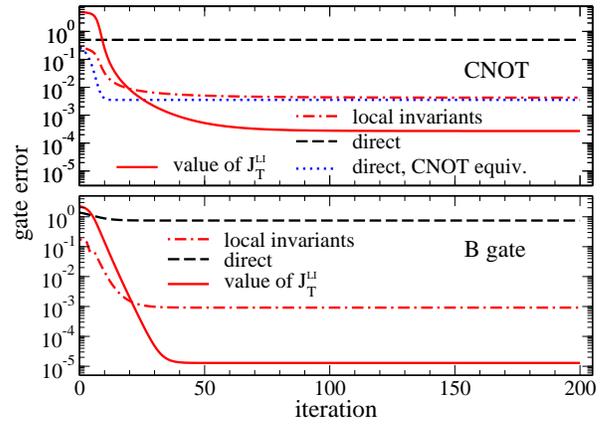}
  \caption{(Color online) 
    Gate error vs iteration step for direct (black dashed lines) and
    local invariant (red dashed-dotted lines) optimization of the 
    CNOT and B gates with time evolution generated by
    Eq.~\eqref{eq:Hnew}. 
    The blue dotted line shows the gate error 
    for direct optimization of a specific instance of the
    local equivalence class [CNOT] (see text). 
    Red solid curves display the value of the local invariants
    functional $J_T^{LI}$ (for 
    the direct optimizations
    the gate error is equal to the value of the
    functional).  
}
  \label{fig:spinspin}
\end{figure}

For a single microwave field, it is straightforward to use the methods
of Section \ref{subsec:U} to determine which non-local equivalence
classes are accessible under time evolution with \eqref{eq:Heff}.
However, multiple fields are employed, as proposed in
Ref. \cite{MicheliNatPhys06}, to allow generation of a broad range of
effective spin-spin Hamiltonians. Whenever the spin-spin interactions
deriving from different fields do not commute, it becomes a
non-trivial task  to determine which two-qubit gates may be
efficiently generated by time evolution under the combined effective
Hamiltonians. Optimization of the
non-local content of the quantum gate reached by time evolution then
provides a useful route to find the acccessible two-qubit gates, an
important task for quantum simulations with these effective spin-spin
Hamiltonians.  We illustrate this with an example of determining which
gates are accessible under irradiation by two microwave fields with
different polarizations.  When one field is pulsed and the other in
continuous wave (cw) modality, the effective Hamiltonian is of
the form 
\begin{equation}
  \label{eq:Hnew}
  \Op H(t) = \Op H_0 + S(t) \Op H_1\,,
\end{equation}
where $\Op H_0$ and $\Op H_1$ do not commute, and $S(t)$ denotes the
envelope (shape) of the pulsed field, $0\le S(t)\le 1$. We 
choose the polarizations to be 
$\alpha^0_0=1/\sqrt{2}$, $\alpha^0_+=1/\sqrt{2}$, and
$\alpha^0_-=0$ for the cw field and $\alpha^1_0=0$,
$\alpha^1_{\pm}=1/\sqrt{2}$ for the pulsed field.
We will first optimize for the CNOT gate. For this case we assume a
rotational transition detuning of 1.2$\,$kHz and a Rabi frequency of 
590$\,$kHz for the cw field and a pulse detuning of 50$\,$kHz. Then
the drift Hamiltonian, in MHz, in the logical basis becomes
\begin{equation}
  \label{eq:H0}
  \Op H_0  = 
  \begin{pmatrix}
    5.711 & 0.324 & 0.324 & 0 \\
    0.324 & -1.840 & 1.054 & 0 \\
    0.324 & 1.054 & 1.840 & 0 \\
    0 & 0 & 0 & -2.030
  \end{pmatrix}\,,
\end{equation}
and the control Hamiltonian, in MHz, is given by
\begin{equation}
  \Op H_1 =  S(t)
  \begin{pmatrix}
    -153.65 & 0 & 0 & 3.906 \\
    0 & 153.65 & 16.085 & 0 \\
    0 & 16.085 & 153.65 & 0 \\
    3.906 & 0 & 0 &  -153.65
  \end{pmatrix}\,.
  \label{eq:H1}
\end{equation}
The peak Rabi frequency of the pulse is $1.81\,$MHz. 
We will then optimize for the B gate.  In this case the polarizations
and therefore the structure of the Hamiltonian matrix 
are the same but the numerical values are slightly
changed since we take different field parameters. Specifically, for the cw field 
we take the rotational transition detuning and
Rabi frequency to be 1.2$\,$kHz and 4.74$\,$MHz,
respectively, and for the pulse field we take
detuning and peak Rabi frequency values of 84$\,$kHz and $1.81\,$MHz,
respectively.  
Figure~\ref{fig:spinspin} shows that direct optimization for CNOT and
B gates is not successful,  
failing to find any high quality solution after a large number of
iterations (dashed black lines). Thus with the structure deriving from
this combination of microwave fields and polarizations, the
Hamiltonian cannot generate the unitary 
transformations corresponding to the B gate and CNOT. In fact, it is 
not evident which gates from which equivalence classes can be realized 
from simply inspecting the Hamiltonian.

In contrast to this lack of success with direct optimization, local
invariant optimization of $J = J_T^{LI} + g_a$ can be successfully
used to find the time-dependent envelope $S(t)$ such that the 
microwave fields implement gates that are locally 
equivalent to the B gate \cite{ZhangPRL04} and to CNOT.
The errors for the $J_T^{LI}$ optimized gates are shown as a function
of iteration number in Fig.~\ref{fig:spinspin} and compared with the
results of direct optimization.  
Progress of optimization in the Weyl chamber is illustrated also
in Fig.~\ref{fig:weyl}. 
The errors for the $J_T^{LI}$ optimized gates in Fig.~\ref{fig:spinspin} 
are seen to be of order $10^{-3}$ (red dashed-dotted line).
The result of standard direct optimization using
$J_T^{D}$ is in stark contrast to this, essentially  
failing to find a solution, as evidenced by the gate error remaining of 
order unity after 200 iterations (dashed black lines). 
This dramatic difference reflects the fact that the  
Hamiltonian  $\Op H_{eff}$, when realized for the field combination of
Eq. (\ref{eq:Hnew}), cannot generate an evolution  
that directly yields either CNOT or B gates. However, 
the successful local invariant optimization 
{\em can} subsequently be used as input for a direct
optimization of CNOT, as follows.
Inspecting the solution
for the local equivalence class [CNOT] that is obtained from optimizing $J_T^{LI}$ 
shows that in this case the optimal unitary transformation is a sum of 
$\Op U_{d}=-\frac{1}{\sqrt{2}}\Fkt{diag}(1-i,1+i,1+i,1-i)$, which is
also in [CNOT] and a smaller  
$\Op U_{od}$ with all matrix elements except  
$U_{23}$, $U_{32}$ zero.
This provides motivation to specify $\Op U_d$ as a target for direct optimization using
$J_T^{D}$.  We find that this leads to a solution of similar quality 
as using $J_T^{LI}$ (dotted blue versus dot-dashed red line in
Fig.~\ref{fig:spinspin}).
In contrast, the optimization of the local equivalence class [B] does
not appear to result in an analogous dominant unitary within its
equivalence class. Thus it is not 
possible to "guess" which representative of [B] might be implemented 
directly.  
This example illustrates how optimization of the locally invariant 
functional can be used to determine what gate operations are
achievable, given a possibly intricate Hamiltonian.

\subsection{Rydberg gate with trapped neutral atoms}
\label{subsec:Rydberg}
An application to time evolution occurring in a  Hilbert
space much larger than that of the quantum register is given by 
qubits encoded in $^{87}$Rb atoms trapped by optical tweezers
\cite{GaetanNatPhys09}. A  
non-local gate is implemented by simultaneous excitation to a Rydberg state using a 
near-resonant  two-photon transition \cite{JakschPRL00}. 

In the experiment of Ref.~\cite{GaetanNatPhys09}, the qubit states are taken
to be $|0\rangle=|5s_{1/2},F=2,M_F=2\rangle$,  
$|1\rangle=|5s_{1/2},F=1,M_F=1\rangle$, the Rydberg state
$|r\rangle=|58d_{3/2},F=3,M_F=3\rangle$, and the intermediate state
for the two-photon transition
$|i\rangle=|5p_{1/2},F=2,M_F=2\rangle$.
In the rotating-wave approximation, the Hamiltonian for a single
trapped atom  reads
\begin{widetext}
\begin{eqnarray}
  \label{eq:Hsingle}
  \Op{H}_j^{(1)}(t) &=& |0\rangle\langle 0| \otimes
  \left(\Op{T} + V_{trap}(\Op{x}_j)\right) +
  |1\rangle\langle 1| \otimes
  \left(\Op{T} + V_{trap}(\Op{x}_j)\right) \nonumber 
  + |i\rangle\langle i| \otimes 
  \left(\Op{T} + \frac{\delta_R}{2}\right)
  + |r\rangle\langle r| \otimes
  \left(\Op{T} +    \frac{\delta_B}{2}\right) \\
  &&+ \frac{\Omega_R(t)}{2} \left( |0\rangle\langle i| + |i\rangle\langle
    0 |\right) \otimes \openone_{\Op{x}_j}
  + \frac{\Omega_B(t)}{2} \left( |i\rangle\langle r| + |r\rangle\langle i
    |\right) \otimes \openone_{\Op{x}_j}\,.   
\end{eqnarray}
\end{widetext}
Here, $\Op{x}_j$ denotes the position operator of atom $j$, 
$\Op{T}$
the kinetic energy operator, and $\Omega_i(t)$ the
time-dependent Rabi frequencies of the red and blue lasers
($\omega_R=795\,$nm and $\omega_B=474\,$nm, respectively).
The maximum Rabi frequencies are taken to be 
$\Omega_{i,0}=2\pi\cdot 260\,$MHz, i.e., $\Omega_{R,0}$ is equal to
and $\Omega_{B,0}$ larger by a factor of 10 than those of~\cite{GaetanNatPhys09}.
$\Omega_i(t)$ is parametrized as 
\[
\Omega_i(t)=\Omega_{i,0}(\tanh\eps_i(t)+1)/2\,\in\,[0,\Omega_{i,0}]\,,
\]
with $\eps_i(t)$ determined by optimal control.
The detuning $\delta_R$ of the red laser is chosen to be
$\delta_R=2\pi\cdot 600\,$MHz, slightly larger than in~\cite{GaetanNatPhys09}.  
The two-photon detuning from the
Rydberg level is given in terms of the Stark shift,
$\delta_B=(\Omega_{B,0}^2-\Omega_{R,0}^2)/(4\delta_R)=0$. 

The total two-atom Hamiltonian includes
the long-range interaction when both atoms are in the Rydberg state,
\begin{eqnarray}
  \label{eq:H_two}
  \Op{H}^{(2)}(t) &=& \Op{H}^{(1)}_1(t) \otimes \openone_{4,2} \otimes
  \openone_{\Op{x}_2} + \openone_{4,1} \otimes \Op{H}^{(1)}_2(t) \otimes 
  \openone_{\Op{x}_1}  \nonumber \\
  &&+ \vert rr\rangle\langle rr\vert 
     \otimes \frac{u_0}{\Op{r}^3}\,,
\end{eqnarray}
with $\Op{r}=|\Op{x}_1-\Op{x}_2|$ the inter-atomic distance. 
$u_0$ is chosen to reproduce the estimated interaction energy
of 50$\,$MHz at $r_0=4\,\mu$m ($u_0=3.284\cdot 10^{6}\,$E$_h$a$_0^3$). 
With this interaction, the atoms need to spend a minimum of
10$\,$ns in $|rr\rangle$ to pick up a non-local phase of $\pi$.
In all other internal states, the interaction between the atoms at a separation
of $4\,\mu$m is negligible. 
We approximate the optical tweezers trap
by harmonic potentials and integrate over the center-of-mass motion.
The trap of width, $\sigma = 0.75\,\mu$m, and depth,
$V_\mathrm{min}=-4.5\,k_B\,$mK, is slightly stronger than 
in Ref.~\cite{GaetanNatPhys09}.

The Hamiltonian, $\Op{H}^{(2)}(t)$, is represented on an
equidistant Fourier grid extending for $\pm 0.3\,\mu$m around $r_0$.
In order to evaluate Eq.~(\ref{eq:field}), the Schr\"odinger equation is
solved simultaneously for initial states $|ij\rangle\cdot
\varphi_0(r)$ ($i,j=0,1$) with $\varphi_0(r)$ the ground state of
the trap, using a Chebychev propagator.

\begin{figure}[bt]
  \centering
  \includegraphics[width=0.95\linewidth]{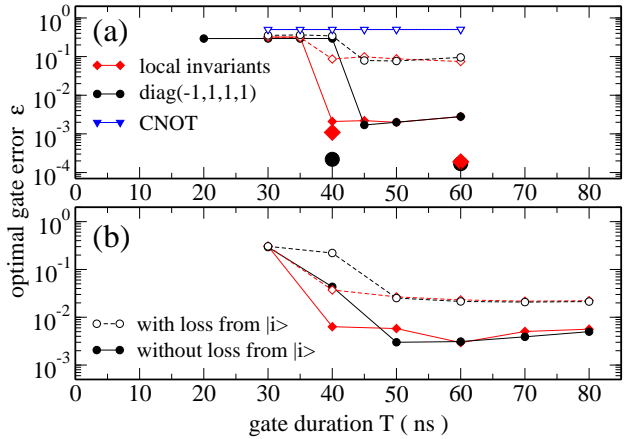}  
  \caption{(Color online)
    Optimal gate error as a function of the gate duration
    for Rydberg gate
    optimization based on local equivalence classes ($J_T^{LI}$) and on
    direct optimization ($J_T^{D}$) of two specific two-qubit gates
    ($\pi$-phase~$\equiv$ diag$(-1,1,1,1)$ and
    CNOT)  
    without (a) and with (b) an additional constraint suppressing
    population in the intermediate state $|i\rangle=|5p_{1/2}\rangle$. 
    Dashed lines with open symbols (solid lines with filled symbols):
    simulation including (neglecting) spontaneous 
    emission from $|i\rangle$. Large filled symbols: calculations with 
    trapping potential kept on during the gate.
  }
  \label{fig:error}
\end{figure}

The errors from the optimal gates are shown in 
Figure~\ref{fig:error} as a function of the corresponding optimal gate duration $T$ and
illustrate several key results. First, the large error resulting from
direct optimization of CNOT (blue triangles) shows that 
$ \Op{H}^{(2)}(t)$
cannot generate CNOT directly. 
Second, the minimum errors of the Rydberg gate 
are comparable for optimization with $J_T^{D}$ and $J_T^{LI}$
(with a slight advantage for the functional based on the local
invariants~\footnote{%
  For local-invariants optimization the gate duration needs to be
  augmented by the duration of the additional one-qubit operations before and after
  the non-local gate. However the one-qubit
  operations are at least one order of magnitude faster than the
  non-local gates.
}) and both of these values 
reflect the quantum speed limit~\cite{TommasoPRL09}. 
Third, if spontaneous emission from $|i\rangle$ is neglected (filled
symbols),  
the minimum gate duration of $\sim 40\,$ns to achieve high fidelities is
determined primarily by the interaction strength in the Rydberg
state.
Motion of the atoms in the trap 
leads to larger minimum gate errors for increasing gate duration.  
Gate errors close to the
fault tolerance threshold of $10^{-4}$ are obtained only when the trapping
potential is kept on during the gate (large filled symbols in
Fig.~\ref{fig:error}(a)).

The main limiting factor for a 
high-fidelity implementation of the Rydberg gate using this particular
near-resonant two-photon transition is 
due to spontaneous emission from the intermediate state $|i\rangle$
as is evident from Fig.~\ref{fig:error} (open symbols).  
Imposing an additional constraint suppressing population
of $|i\rangle$, cf. Eq.~\eqref{eq:g_b}, 
leads to improved
solutions but leaves the minimum errors still two orders of
magnitude above  the fault tolerant threshold (Fig.~\ref{fig:error}(b)). 
Near resonant intermediate states should therefore be avoided by
suitable choice of Rydberg and intermediate states, for given laser
frequencies.

\begin{figure}[bt]
  \centering
  \includegraphics[width=0.9\linewidth]{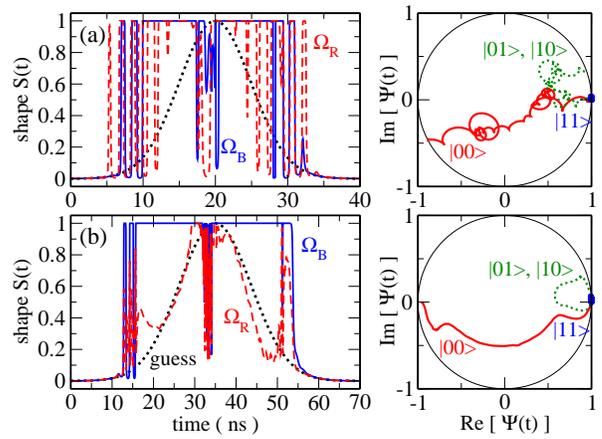}  
  \caption{(Color online)
    Left: optimal Rabi frequencies without (a) and with (b) spontaneous
    emission from $|i\rangle$.  Black dotted line represents the
    initial "guess" pulse form.  Right: corresponding dynamics
    of the two-qubit basis states in the complex plane.
  } 
  \label{fig:dyn}
\end{figure}

Sample Rabi frequencies for optimized pulses and the corresponding
two-qubit state dynamics are shown in Fig.~\ref{fig:dyn}.
Without spontaneous emission, the minimum gate duration for which a
high-fidelity implementation of CNOT can be obtained is 40$\,$ns. 
The optimal pulses consist of a sequence of fast switches pumping
population in a ladder-like fashion. This is reflected by the small
circles of $|01\rangle$, $|10\rangle$ states in the complex plane,
indicating clearly non-adiabatic dynamics. 
The time between switches corresponds to the duration of a $\pi$-pulse
at the maximally allowed Rabi frequencies
$\Omega_{i,0}$. Correspondingly, the spectra (not shown) consist of
modulated sidebands spaced by $\Omega_{i,0}$. Experimentally, 
switches of the field amplitude on a nanosecond timescale can be
generated using standard electronics (electro- or
acousto-optical modulators).
When spontaneous emission from the intermediate state is taken into
account in the optimization using the additional constraint
$g_b$, the smallest gate errors  achievable for the given detunings and
maximum Rabi frequencies are of a few percent. This requires gate
durations of 50$\,$ns or more, which is \textit{larger} than the gate durations
without spontaneous emission. While surprising at first glance, it
reflects the fact that the only way to avoid populating $|i\rangle$ is by
adiabatic passage, and adiabaticity requires sufficient 
time. Correspondingly, the optimal Rabi frequencies in
Fig.~\ref{fig:dyn}(b) show a double STIRAP-like behavior
\cite{BergmannRMP98}: In the first half of the time interval, 
the blue pulse connecting $|i\rangle$ and
$|r\rangle$ takes the role of the Stokes laser and precedes the red
pulse connecting $|0\rangle$ and $|i\rangle$ which corresponds to the
pump laser. In the second half of the time interval, these roles are
reversed. The optimal entangling strategies in the two examples are
thus based on well-known, robust and feasible control schemes --
population transfer to and from the Rydberg state via $\pi$-pulses or
STIRAP. The additional twist that is afforded by optimal control
is speed-up, i.e., implementation of the shortest possible gate
duration. This comes at a comparatively low price -- additional
modulations in the optimal fields on a nanosecond timescale 
translating into spectra whose bandwidth never exceeds a few GHz.
It should therefore be comparatively straightforward to implement
these optimal pulses in an experiment.  

Both with and without spontaneous emission, the qubit phase dynamics
(right panel of Fig.~\ref{fig:dyn}) show achievement 
of the desired non-local 
phase, $\chi = \phi_{00} - \phi_{01} - \phi_{10} + \phi_{11}$, 
with complete freedom in the local phases. This confirms the fact that
$J_T^{LI}$ imposes fewer constraints than direct
optimization.
Since spontaneous emission from intermediate states can be eliminated
by suitable choice of atomic states and exciting lasers, it is evident
that this non-local optimal control allows a high-fidelity implementation
of the Rydberg gate to 
be achieved even for a setting where 
the blockade regime is not reached and when additional 
entanglement between qubit
and external degrees of freedom is allowed. 

\section{Summary and Conclusions}

By constructing a new optimization functional based on the local
invariants of two-qubit operations, we have shown how to extend
optimal control theory to take specific requirements of
quantum information applications into account. 
In particular, we have developed an
automated way to determine for a given physical system whether a
desired non-local content can be realized, and if so which two-qubit
operation in that desired local equivalence class can best be
realized.
We have illustrated the power of this approach with two examples
relevant to quantum simulations and quantum computation.  The first
example addressed the time dependent control of effective spin
Hamiltonians generated by trapped dipolar molecules. It showed that
optimization of non-local content of two-qubit operations by local
invariant optimization allows determination of which gates may be
reached from complex Hamiltonians with non-commutative time
dependence.  The second example addressed the performance of a Rydberg
gate with trapped atoms. It demonstrated that use of optimal control theory
with the local invariant functional yields a faster gate than direct
optimization, in the presence of both coupling between internal and
external degrees of freedom and spontaneous emission. 

The new optimization functional $J_T^{LI}$ can easily be adapted to target
all perfect entanglers, using the geometric definition of these
derived in \cite{ZhangPRA03}. 
A related question of optimizing multi-partite entanglement was
recently addressed using time-local control theory
\cite{PlatzerPRL10}. 
Local invariant optimization is expected to display its full potential in the
presence of decoherence, particularly when distinct two-qubit gates in the same local
equivalence class are differently affected by the decoherence. 
The required extension to open quantum systems is the
subject of future work. 

\begin{acknowledgments}
  We enjoyed hospitality of the KITP during the 2009 Quantum
  Control of Light and Matter program (NSF-KITP-11-047).
  Financial support from the EC projects AQUTE, PICC, and DIAMANT, the BMBF
  network QuOReP,  Science Foundation Ireland
  and the Deutsche
  Forschungsgemeinschaft 
  is gratefully acknowledged. 
  We thank the bwGRiD project (www.bw-grid.de)
  for computational resources.
\end{acknowledgments}

\appendix

\section{Explicit form of the local invariants functional}
\label{app:J}

Using the definitions of $\Op m$ and the local
invariants, Eqs.~\eqref{eq:m} and \eqref{eq:gs}, 
the functional is expressed in terms of the
states,
\begin{displaymath}
  J_{T}^{LI} = f_{1}^{2}+f_{2}^{2}+f_{3}^{2}+f_{4}^{2} + 1 -
  \frac{1}{N} \Fkt{Tr}\left\{\Op U_{T,N} \Op U_{T,N}^+\right\}
\end{displaymath}
with
\begin{eqnarray*}
  f_{1} & = & \mathfrak{Re}\left[a_{0}\Fkt{det}\left\{\Op U_{T,N}\right\}\right] 
   -\frac{1}{16}\sum_{k,l}\bigg[
    \vec{\alpha}_{k}^{2}\vec{\alpha}_{l}^{2}
    +\vec{\beta}_{k}^{2}\vec{\beta}_{l}^{2}\\
   && -2\vec{\alpha}_{k}^{2}\vec{\beta}_{l}^{2} 
   -4\left(\vec{\alpha}_{k}\cdot\vec{\beta}_{k}\right)
    \left(\vec{\alpha}_{l}\cdot\vec{\beta}_{l}\right) \bigg] \\
  f_{2} & = & \mathfrak{Im}\left[a_{0}\Fkt{det}\left\{\Op U_{T,N}\right\}\right]
   -\frac{1}{16}\sum_{k,l}\bigg[
    4\vec{\alpha}_{k}^{2}\left(\vec{\alpha}_{l}\cdot\vec{\beta}_{l}\right)\\
    &&-4\vec{\beta}_{k}^{2}\left(\vec{\alpha}_{l}\cdot\vec{\beta}_{l}\right)\bigg]\\
  f_{3} & = & \mathfrak{Re}\left[b_{0}\Fkt{det}\left\{\Op U_{T,N}\right\}\right]
  -\frac{1}{4}\sum_{k,l}\bigg[
    \vec{\alpha}_{k}^{2}\vec{\alpha}_{l}^{2}+\vec{\beta}_{k}^{2}\vec{\beta}_{l}^{2}
    -2\vec{\alpha}_{k}^{2}\vec{\beta}_{l}^{2}\\
   && -4\left(\vec{\alpha}_{k}\cdot\vec{\beta}_{k}\right)
    \left(\vec{\alpha}_{l}\cdot\vec{\beta}_{l}\right)
    -\left(\vec{\alpha}_{k}\cdot\vec{\alpha}_{l}\right)^{2}
    -\left(\vec{\beta}_{k}\cdot\vec{\beta}_{l}\right)^{2}\\
    &  & +2\left(\vec{\alpha}_{k}\cdot\vec{\alpha}_{l}\right)
    \left(\vec{\beta}_{k}\cdot\vec{\beta}_{l}\right)
    +4\left(\vec{\alpha}_{k}\cdot\vec{\alpha}_{l}\right)
    \left(\vec{\beta}_{k}\cdot\vec{\beta}_{l}\right)\bigg]\\
  f_{4} & = &  \mathfrak{Im}\left[b_{0}\Fkt{det}\left\{\Op U_{T,N}\right\}\right] 
  -\frac{1}{4}\sum_{k,l}\bigg[4\vec{\alpha}_{k}^{2}
  \left(\vec{\alpha}_{l}\cdot\vec{\beta}_{l}\right)\\
  &&-4\vec{\beta}_{k}^{2}\left(\vec{\alpha}_{l}\cdot\vec{\beta}_{l}\right)
  -4\left(\vec{\alpha}_{k}\cdot\vec{\alpha}_{l}\right)
  \left(\vec{\alpha}_{k}\cdot\vec{\beta}_{l}\right)\\
  &&+4\left(\vec{\beta}_{k}\cdot\vec{\beta}_{l}\right)
  \left(\vec{\alpha}_{k}\cdot\vec{\beta}_{l}\right)\bigg]\,,
\end{eqnarray*}
where the sum runs over the $N$ logical basis states.
The constants $a_0$, $b_0$ are obtained by calculating $\Op m$,
cf. Eq.~\eqref{eq:m}, for  the target gate $\Op{O}$,
\begin{eqnarray*}
a_{0} & = & \frac{\Fkt{Tr}^{2}\left\{\Op{m}_{O}\right\}}{16\Fkt{det}\{\Op{O}\}}\\
b_{0} & = &
\frac{\left[\Fkt{Tr}^{2}\left\{\Op{m}_{O}\right\}
-\Fkt{Tr}\left\{\Op{m}_{O}^{2}\right\}\right]}{4\Fkt{det}\{\Op{O}\}}\,,
\end{eqnarray*}
and $\vec{\alpha}_k$ ($\vec{\beta}_k$) is the vector containing all real
(imaginary) parts of the expansion coefficients of the state
$|\vphi_k\rangle$ with respect to an orthonormal basis, $\{|m\rangle\}$, spanning
the complete Hilbert space,
\begin{eqnarray*}
  (\alpha_k)_m &=& \mathfrak{Re}\left[\langle m|\vphi_k(t)\rangle\right]
  \,,\, m=1,\ldots , \Fkt{dim}(\mathcal{H}) \\
  (\beta_k)_m &=& \mathfrak{Im}\left[\langle m|\vphi_k(t)\rangle\right]
  \,,\, m=1,\ldots , \Fkt{dim}(\mathcal{H})\,.
\end{eqnarray*}
Note that $J_T^{LI}$ is a polynomial of 8th degree
in the states. 

\section{Outline of the optimization algorithm}
\label{app:OCT}

The algorithm is determined by the functional 
$J_T$, additional time-dependent constraints $g_a$, $g_b$ and the
equations of motion~\cite{Reich10} and 
given by the following set of equations:
\begin{enumerate}
\item Forward propagation to obtain the new states
$|\vphi_k^{(i+1)}(t)\rangle$, 
\begin{equation}
  \label{eq:forward}
\frac{d}{dt}|\vphi_k^{(i+1)}(t)\rangle =
-\frac{i}{\hbar}\Op{H}[\epsilon^{(i+1)}] |\vphi_k^{(i+1)}(t)\rangle   \,.
\end{equation}
The initial states are given by the basis expansion of the time
evolution operator, i.e., the logical basis states in our case.
\item Backward propagation to obtain the adjoint states
$|\chi_k^{(i)}(t)\rangle$, 
containing an inhomogeneity if $g_b\neq 0$,
\begin{equation}
  \label{eq:backward}
\frac{d}{dt}|\chi_k^{(i)}(t)\rangle =
-\frac{i}{\hbar}\Op{H}[\epsilon^{(i)}]
|\chi_k^{(i)}(t)\rangle + |\eta\rangle  
\end{equation}
with the 'initial' condition at time $t=T$ determined by $J_T$,
\begin{equation}
  \label{eq:chi_ini}
  |\chi_k(T)\rangle = \nabla_{\langle\vphi|} J_T\,,  
\end{equation}
and the inhomogeneity
\begin{equation}
  \label{eq:eta}
  |\eta\rangle =  \nabla_{\langle\vphi|} g_b \,.
\end{equation}
\item The equation to determine the new field
from $|\vphi_k^{(i+1)}(t)\rangle$ and
$|\chi_k^{(i)}(t)\rangle$, Eq.~\eqref{eq:field}, 
with  $\sigma(t)$ given by
\begin{equation}
  \label{eq:sigma}
  \sigma(t) = C(T-t)-A\,.
\end{equation}
The constants $A$ and $C$ are parameters
of the algorithm that can be estimated analytically (based on a worst
case scenario) or numerically~\cite{Reich10}. For the local invariants
functional, $J_T^{LI}$, the analytical estimate yields 
$A=90$ for Hamiltonian \eqref{eq:Heff} and $A=580$ for Hamiltonian
\eqref{eq:H_two}, while numerically $A =5$ and $A=20$, respectively,
turned out to be sufficient. 
$C=0$ for $g_b=0$ and $C\le -\lambda_b/NT$ for
$g_b\neq 0$.
\end{enumerate}
Compared to a linear version of the Krotov
algorithm~\cite{PalaoPRA03}, the additional effort consists only in 
storing the forward propagated states from the previous iteration, 
$|\vphi_k^{(i)}(t)\rangle$, and calculating 
$|\Delta\vphi_k(t)\rangle$ and $\sigma(t)$.


\begin{thebibliography}{19}
\expandafter\ifx\csname natexlab\endcsname\relax\def\natexlab#1{#1}\fi
\expandafter\ifx\csname bibnamefont\endcsname\relax
  \def\bibnamefont#1{#1}\fi
\expandafter\ifx\csname bibfnamefont\endcsname\relax
  \def\bibfnamefont#1{#1}\fi
\expandafter\ifx\csname citenamefont\endcsname\relax
  \def\citenamefont#1{#1}\fi
\expandafter\ifx\csname url\endcsname\relax
  \def\url#1{\texttt{#1}}\fi
\expandafter\ifx\csname urlprefix\endcsname\relax\def\urlprefix{URL }\fi
\providecommand{\bibinfo}[2]{#2}
\providecommand{\eprint}[2][]{\url{#2}}

\bibitem[{\citenamefont{Nielsen and Chuang}(2000)}]{NielsenChuang}
\bibinfo{author}{\bibfnamefont{M.}~\bibnamefont{Nielsen}} \bibnamefont{and}
  \bibinfo{author}{\bibfnamefont{I.~L.} \bibnamefont{Chuang}},
  \emph{\bibinfo{title}{Quantum Computation and Quantum Information}}
  (\bibinfo{publisher}{Cambridge University Press}, \bibinfo{year}{2000}).

\bibitem[{\citenamefont{Palao and Kosloff}(2003)}]{PalaoPRA03}
\bibinfo{author}{\bibfnamefont{J.~P.} \bibnamefont{Palao}} \bibnamefont{and}
  \bibinfo{author}{\bibfnamefont{R.}~\bibnamefont{Kosloff}},
  \bibinfo{journal}{Phys. Rev. A} \textbf{\bibinfo{volume}{68}},
  \bibinfo{pages}{062308} (\bibinfo{year}{2003}).

\bibitem[{\citenamefont{Zhang et~al.}(2003)\citenamefont{Zhang, Vala, Sastry,
  and Whaley}}]{ZhangPRA03}
\bibinfo{author}{\bibfnamefont{J.}~\bibnamefont{Zhang}},
  \bibinfo{author}{\bibfnamefont{J.}~\bibnamefont{Vala}},
  \bibinfo{author}{\bibfnamefont{S.}~\bibnamefont{Sastry}}, \bibnamefont{and}
  \bibinfo{author}{\bibfnamefont{K.~B.} \bibnamefont{Whaley}},
  \bibinfo{journal}{Phys. Rev. A} \textbf{\bibinfo{volume}{67}},
  \bibinfo{pages}{042313} (\bibinfo{year}{2003}).

\bibitem[{\citenamefont{Micheli et~al.}(2006)\citenamefont{Micheli, Brennen,
  and Zoller}}]{MicheliNatPhys06}
\bibinfo{author}{\bibfnamefont{A.}~\bibnamefont{Micheli}},
  \bibinfo{author}{\bibfnamefont{G.~K.} \bibnamefont{Brennen}},
  \bibnamefont{and} \bibinfo{author}{\bibfnamefont{P.}~\bibnamefont{Zoller}},
  \bibinfo{journal}{Nature Phys.} \textbf{\bibinfo{volume}{2}},
  \bibinfo{pages}{341 } (\bibinfo{year}{2006}).

\bibitem[{\citenamefont{Jaksch et~al.}(2000)\citenamefont{Jaksch, Cirac,
  Zoller, Rolston, C\^ot\'e, and Lukin}}]{JakschPRL00}
\bibinfo{author}{\bibfnamefont{D.}~\bibnamefont{Jaksch}},
  \bibinfo{author}{\bibfnamefont{J.~I.} \bibnamefont{Cirac}},
  \bibinfo{author}{\bibfnamefont{P.}~\bibnamefont{Zoller}},
  \bibinfo{author}{\bibfnamefont{S.~L.} \bibnamefont{Rolston}},
  \bibinfo{author}{\bibfnamefont{R.}~\bibnamefont{C\^ot\'e}}, \bibnamefont{and}
  \bibinfo{author}{\bibfnamefont{M.~D.} \bibnamefont{Lukin}},
  \bibinfo{journal}{Phys. Rev. Lett.} \textbf{\bibinfo{volume}{85}},
  \bibinfo{pages}{2208} (\bibinfo{year}{2000}).

\bibitem[{\citenamefont{Palao et~al.}(2008)\citenamefont{Palao, Kosloff, and
  Koch}}]{PalaoPRA08}
\bibinfo{author}{\bibfnamefont{J.~P.} \bibnamefont{Palao}},
  \bibinfo{author}{\bibfnamefont{R.}~\bibnamefont{Kosloff}}, \bibnamefont{and}
  \bibinfo{author}{\bibfnamefont{C.~P.} \bibnamefont{Koch}},
  \bibinfo{journal}{Phys. Rev. A} \textbf{\bibinfo{volume}{77}},
  \bibinfo{pages}{063412} (\bibinfo{year}{2008}).

\bibitem[{\citenamefont{Konnov and Krotov}(1999)}]{Konnov99}
\bibinfo{author}{\bibfnamefont{A.}~\bibnamefont{Konnov}} \bibnamefont{and}
  \bibinfo{author}{\bibfnamefont{V.}~\bibnamefont{Krotov}},
  \bibinfo{journal}{Automation and Remote Control}
  \textbf{\bibinfo{volume}{60}}, \bibinfo{pages}{1427} (\bibinfo{year}{1999}).

\bibitem[{\citenamefont{Sklarz and Tannor}(2002)}]{SklarzPRA02}
\bibinfo{author}{\bibfnamefont{S.~E.} \bibnamefont{Sklarz}} \bibnamefont{and}
  \bibinfo{author}{\bibfnamefont{D.~J.} \bibnamefont{Tannor}},
  \bibinfo{journal}{Phys. Rev. A} \textbf{\bibinfo{volume}{66}},
  \bibinfo{pages}{053619} (\bibinfo{year}{2002}).

\bibitem[{\citenamefont{Zhu et~al.}(1998)\citenamefont{Zhu, Botina, and
  Rabitz}}]{ZhuJCP98}
\bibinfo{author}{\bibfnamefont{W.}~\bibnamefont{Zhu}},
  \bibinfo{author}{\bibfnamefont{J.}~\bibnamefont{Botina}}, \bibnamefont{and}
  \bibinfo{author}{\bibfnamefont{H.}~\bibnamefont{Rabitz}},
  \bibinfo{journal}{J. Chem. Phys.} \textbf{\bibinfo{volume}{108}},
  \bibinfo{pages}{1953} (\bibinfo{year}{1998}).

\bibitem[{\citenamefont{Maday and Turinici}(2003)}]{MadayJCP03}
\bibinfo{author}{\bibfnamefont{Y.}~\bibnamefont{Maday}} \bibnamefont{and}
  \bibinfo{author}{\bibfnamefont{G.}~\bibnamefont{Turinici}},
  \bibinfo{journal}{J. Chem. Phys.} \textbf{\bibinfo{volume}{118}},
  \bibinfo{pages}{8191} (\bibinfo{year}{2003}).

\bibitem[{\citenamefont{Goerz et~al.}(2011)\citenamefont{Goerz, Calarco, and
  Koch}}]{Goerz}
\bibinfo{author}{\bibfnamefont{M.~H.} \bibnamefont{Goerz}},
  \bibinfo{author}{\bibfnamefont{T.}~\bibnamefont{Calarco}}, \bibnamefont{and}
  \bibinfo{author}{\bibfnamefont{C.~P.} \bibnamefont{Koch}},
  \bibinfo{journal}{J. Phys. B} \textbf{\bibinfo{volume}{44}},
  \bibinfo{pages}{150411} (\bibinfo{year}{2011}), \bibinfo{note}{special issue
  on quantum control theory for coherence and information dynamics}.

\bibitem[{\citenamefont{Reich et~al.}()\citenamefont{Reich, Ndong, and
  Koch}}]{Reich10}
\bibinfo{author}{\bibfnamefont{D.~M.} \bibnamefont{Reich}},
  \bibinfo{author}{\bibfnamefont{M.}~\bibnamefont{Ndong}}, \bibnamefont{and}
  \bibinfo{author}{\bibfnamefont{C.~P.} \bibnamefont{Koch}},
  \eprint{1008.5126}.

\bibitem[{\citenamefont{Helgason}(1978)}]{Helgason78}
\bibinfo{author}{\bibfnamefont{S.}~\bibnamefont{Helgason}},
  \emph{\bibinfo{title}{Differential geometry, {L}ie groups, and symmetric
  spaces}} (\bibinfo{publisher}{Academic, New York}, \bibinfo{year}{1978}).

\bibitem[{\citenamefont{Makhlin}(2002)}]{MakhlinQIP02}
\bibinfo{author}{\bibfnamefont{Y.}~\bibnamefont{Makhlin}},
  \bibinfo{journal}{Quant. Inf. Proc.} \textbf{\bibinfo{volume}{1}},
  \bibinfo{pages}{243} (\bibinfo{year}{2002}).

\bibitem[{\citenamefont{Zhang et~al.}(2004)\citenamefont{Zhang, Vala, Sastry,
  and Whaley}}]{ZhangPRL04}
\bibinfo{author}{\bibfnamefont{J.}~\bibnamefont{Zhang}},
  \bibinfo{author}{\bibfnamefont{J.}~\bibnamefont{Vala}},
  \bibinfo{author}{\bibfnamefont{S.}~\bibnamefont{Sastry}}, \bibnamefont{and}
  \bibinfo{author}{\bibfnamefont{B.}~\bibnamefont{Whaley}},
  \bibinfo{journal}{Phys. Rev. Lett.} \textbf{\bibinfo{volume}{93}},
  \bibinfo{pages}{020502} (\bibinfo{year}{2004}).

\bibitem[{\citenamefont{Ga\"{e}tan et~al.}(2009)\citenamefont{Ga\"{e}tan,
  Miroshnychenko, Wilk, Chotia, Vitaeu, Comparat, Pillet, Browaeys, and
  Grangier}}]{GaetanNatPhys09}
\bibinfo{author}{\bibfnamefont{A.}~\bibnamefont{Ga\"{e}tan}},
  \bibinfo{author}{\bibfnamefont{Y.}~\bibnamefont{Miroshnychenko}},
  \bibinfo{author}{\bibfnamefont{T.}~\bibnamefont{Wilk}},
  \bibinfo{author}{\bibfnamefont{A.}~\bibnamefont{Chotia}},
  \bibinfo{author}{\bibfnamefont{M.}~\bibnamefont{Vitaeu}},
  \bibinfo{author}{\bibfnamefont{D.}~\bibnamefont{Comparat}},
  \bibinfo{author}{\bibfnamefont{P.}~\bibnamefont{Pillet}},
  \bibinfo{author}{\bibfnamefont{A.}~\bibnamefont{Browaeys}}, \bibnamefont{and}
  \bibinfo{author}{\bibfnamefont{P.}~\bibnamefont{Grangier}},
  \bibinfo{journal}{Nature Phys.} \textbf{\bibinfo{volume}{5}},
  \bibinfo{pages}{115} (\bibinfo{year}{2009}).

\bibitem[{\citenamefont{Caneva et~al.}(2009)\citenamefont{Caneva, Murphy,
  Calarco, Fazio, Montangero, Giovannetti, and Santoro}}]{TommasoPRL09}
\bibinfo{author}{\bibfnamefont{T.}~\bibnamefont{Caneva}},
  \bibinfo{author}{\bibfnamefont{M.}~\bibnamefont{Murphy}},
  \bibinfo{author}{\bibfnamefont{T.}~\bibnamefont{Calarco}},
  \bibinfo{author}{\bibfnamefont{R.}~\bibnamefont{Fazio}},
  \bibinfo{author}{\bibfnamefont{S.}~\bibnamefont{Montangero}},
  \bibinfo{author}{\bibfnamefont{V.}~\bibnamefont{Giovannetti}},
  \bibnamefont{and} \bibinfo{author}{\bibfnamefont{G.~E.}
  \bibnamefont{Santoro}}, \bibinfo{journal}{Phys. Rev. Lett.}
  \textbf{\bibinfo{volume}{103}}, \bibinfo{pages}{240501}
  (\bibinfo{year}{2009}).

\bibitem[{\citenamefont{Bergmann et~al.}(1998)\citenamefont{Bergmann, Theuer,
  and Shore}}]{BergmannRMP98}
\bibinfo{author}{\bibfnamefont{K.}~\bibnamefont{Bergmann}},
  \bibinfo{author}{\bibfnamefont{H.}~\bibnamefont{Theuer}}, \bibnamefont{and}
  \bibinfo{author}{\bibfnamefont{B.~W.} \bibnamefont{Shore}},
  \bibinfo{journal}{Rev. Mod. Phys.} \textbf{\bibinfo{volume}{70}},
  \bibinfo{pages}{1003} (\bibinfo{year}{1998}).

\bibitem[{\citenamefont{Platzer et~al.}(2010)\citenamefont{Platzer, Mintert,
  and Buchleitner}}]{PlatzerPRL10}
\bibinfo{author}{\bibfnamefont{F.}~\bibnamefont{Platzer}},
  \bibinfo{author}{\bibfnamefont{F.}~\bibnamefont{Mintert}}, \bibnamefont{and}
  \bibinfo{author}{\bibfnamefont{A.}~\bibnamefont{Buchleitner}},
  \bibinfo{journal}{Phys. Rev. Lett.} \textbf{\bibinfo{volume}{105}},
  \bibinfo{pages}{020501} (\bibinfo{year}{2010}).

\end{thebibliography}

\end{document}